\pdfoutput=1
\documentclass[reprint, amsmath,amssymb, aps, prx, longbibliography]{revtex4-2}

\usepackage{footmisc,multirow}
\usepackage{subfigure} 
\usepackage{amsmath,slashed}
\usepackage{upgreek}
\usepackage{graphicx}
\begin{document}


\title{Probing new physics using Rydberg states of atomic hydrogen}

\author{Matthew P. A. Jones}  
\email{m.p.a.jones@durham.ac.uk}
\affiliation{Department of Physics, Joint Quantum Centre Durham-Newcastle, Durham University, South Road, Durham DH1 3LE, England, United Kingdom}

\author{Robert M. Potvliege}
\email{r.m.potvliege@durham.ac.uk}
\affiliation{Department of Physics, Joint Quantum Centre Durham-Newcastle, Durham University, South Road, Durham DH1 3LE, England, United Kingdom}

\author{Michael Spannowsky}
\email{michael.spannowsky@durham.ac.uk}
\affiliation{Department of Physics, Institute of Particle Physics Phenomenology, Durham University, South Road, Durham DH1 3LE, England, United Kingdom}



\begin{abstract}
We consider the role of high-lying Rydberg states of simple atomic systems such as $^1\mathrm{H}$ in setting constraints on physics beyond the Standard Model. 
We obtain highly accurate bound states energies for a hydrogen atom in the presence of an additional force carrier (the energy levels of the Hellmann potential).
These results show that varying the size and shape of the Rydberg state by varying the quantum numbers provides a way to probe the range of new forces. 
By combining these results with the current state-of-the-art QED corrections, we determine a robust global constraint on new physics that includes all current spectroscopic data in hydrogen. Lastly we show that improved measurements that fully exploit modern cooling and trapping methods as well as higher-lying states could lead to a strong, statistically robust global constraint on new physics based 
on laboratory measurements only.

\end{abstract}

\maketitle


\section{Introduction}
\label{sec:intro}
 Detailed measurements of atomic spectra were key to the discovery of quantum mechanics and the development of relativistic quantum electrodynamics (QED). Today, precision atomic spectroscopy underpins the SI system of units, provides the values of some fundamental constants, and enables precise tests of Standard Model calculations. 

Looking for deviations between precise spectroscopic measurements and their Standard Model predictions thus provides a powerful way to set constraints on new physics \cite{Safronova2018}. One powerful approach looks for small effects that break symmetries such as parity (P violation) or time-reversal (T-violation). Alternatively, one can  compare experimental and theoretical transition frequencies. If additional force mediators (bosons) were present that coupled strongly enough to the nucleus and electrons, they would modify the frequency of spectral lines. Thus, by comparing experimentally measured spectra with theory the existence of new so-called fifth forces can be tested down to very small interaction strengths. In recent years extensions of the Standard Model, e.g., modified gravity \cite{Brax:2010gp, Brax:2014zba}, axions \cite{Frugiuele:2016rii, Berengut:2017zuo}, new gauge boson \cite{Jaeckel:2010xx, Jentschura:2018zjv}, have been constrained in this way.  In particular if the force mediator $X$ is light, i.e., below 1 MeV in mass, and couples to partons and electrons, the limits obtained from atomic spectroscopy are many orders of magnitude stronger than from any other laboratory-based experiment, including high-energy collider experiments \cite{Karshenboim:2010cg,Karshenboim:2010ck,Jaeckel:2010xx}.

While modifications of the Standard Model through light bosons are predicted by various models, they arguably receive strong constraints from astrophysical sources \cite{Grifols:1986fc,Raffelt:2012sp, Viaux:2013lha}, e.g. the energy loss from the sun, globular clusters or supernovae. However, the need for independent laboratory-based experiments has been pointed out frequently --- see, e.g., \cite{Masso:2005ym,Jaeckel:2006xm,Jaeckel:2006id,Brax:2007ak}. As an example, a prominent class of light-scalar models  potentially related to modified gravity and dark energy, are chameleons \cite{Khoury:2003aq,Brax:2007ak,Burrage:2014oza}. Chameleons have a mass that depends on the energy density of their environment and thus can avoid being produced in stars, thereby avoiding astrophysical bounds.

One of the main uncertainties in the prediction of spectral lines arises due to the difficulty of solving the Schr{\"o}dinger or Dirac equations for many interacting electrons. Even state-of-the-art calculations for species commonly used in atomic clocks only attain a fractional uncertainty of $\sim 10^{-5}$ \cite{Safronova2008}, which is 14 orders of magnitude lower than the current experimental precision. To circumvent this limitation, it has been proposed to look for new physics using the difference in spectral line positions between isotopes (isotope shifts) \cite{Delaunay:2016brc,Berengut:2017zuo}, rather than by direct comparison with theory. Although promising \cite{Ohayon2019}, the method is limited by the requirement that at least three stable isotopes with two suitable transitions exist for each element. 

An alternative approach is to use light atomic species such as H or He for which full standard model predictions of line positions including QED corrections (Lamb shift) and weak interactions (Z boson exchange) are possible.  Even here however, the complex structure of the nucleus, in particular the details of its charge distribution, limits the achievable accuracy. Spectroscopic data that does not strongly depend on the details of how the nucleus is modelled can thus help to improve the sensitivity on the presence of new forces.

In this paper, we explore how the precision spectroscopy of states with a high principal quantum number $n$ (Rydberg states) might be used to set constraints on physics beyond the Standard Model. 
In principle such states offer several advantages that could be exploited in a search for new physics. Firstly, the overlap of Rydberg wave functions with the nucleus scales as $n^{-3}$, vastly reducing their sensitivity to nuclear effects. The radiative lifetime also scales as $n^{-3}$, meaning that narrow transitions from low-lying atomic states are available that span the UV to NIR wavelength range that amenable to precision laser spectroscopy. The $n^{-2}$ scaling of the energy levels means that for each atomic species a large number of such transitions are available within a narrow spectral range. Lastly, there is a  natural length scale associated with the atomic wave function that scales as $n^2$. As we will show, being able to vary this length scale enables tests which are sensitive to the corresponding length scale associated with any new forces
\cite{Karshenboim:2010cg}.

Here we take hydrogen as a model system in which to explore the use of Rydberg states in searches for new physics. Measurements with a fractional uncertainty of 10 ppt or better are already available for $n$ up to 12. We calculate the (non-relativistic) spectrum of the combination of a hydrogenic Coulomb potential and a Yukawa potential arising from new physics to high accuracy. By combining the resulting energies with previously derived relativistic, QED and hyperfine corrections, we obtain predicted atomic transition frequencies that can be compared directly to experimental data to set a constraint on the strength of a new physics interaction. 
We consider in detail how uncertainties due to the Rydberg constant and the proton charge radius can be reduced or eliminated altogether, and show how a global statistical analysis can be used to derive robust atomic physics constraints. Lastly we develop proposals for future improved tests using atomic Rydberg spectroscopy in atomic hydrogen and other species.

The structure of the paper is as follows. In Section \ref{sec:param} we introduce the simplified model used to parameterize the effect of new physics. Calculations of the effect of new bosons on atomic energy levels are presented in section \ref{sec:NPshifts}. We assess the current experimental reach for new physics (NP) in Section \ref{section:current}. In Section~\ref{sec:improve} we discuss the impact of potential experimental and theoretical improvements on the uncertainty budget and in how far this can result in tighter constraints of new physics. We offer a summary and conclusions in Section~\ref{sec:conc}.

\section{Parametrisation of new physics}
\label{sec:param}
With the discovery of the Higgs boson \cite{Aad:2012tfa,Chatrchyan:2012xdj}, for the first time a seemingly elementary scalar sector  was established in nature. Such a particle would mediate a new short-ranged force, the so-called Higgs boson force \cite{Haber:1978jt}. While the Higgs boson force is very difficult to measure in atom spectroscopy \cite{Delaunay:2016brc}, many extensions of the Standard Model predict elementary scalar or vector particles with a very light mass. Examples include axions \cite{Frugiuele:2016rii,Berengut:2017zuo,Gupta:2019ueh},  modified-gravity models \cite{Brax:2010gp,Brax:2010jk}, millicharged particles \cite{Abel:2008ai, Goodsell:2009xc,Jaeckel:2010ni}, Higgs-portal models \cite{Schabinger:2005ei,Patt:2006fw} and light $Z'$ \cite{Holdom:1985ag,Foot:1991kb}.

To remain as model-independent as possible in parameterizing deformations from the Standard Model (SM), it has become standard practice to express new physics contributions in terms of so-called simplified models \cite{Alves:2011wf}. The idea is to add new degrees of freedom to the Standard Model Lagrangian without asking how such states arise from a UV complete theory. Thus, one can describe the dynamics and phenomenological implications of new degrees of freedom without making further assumptions on the UV theory from which they descend {}\footnote{Although not UV complete, simplified models are effective theories, and as such are consistent up to a certain energy scale. Their range of validity is therefore limited. However, they have the merit of establishing a connection between the parameters of the Yukawa potential of Eq.~(\ref{eq:yukpot}) and the fundamental parameters of a (simplified) Lagrangian. Their incompleteness does not compromise the validity of the Yukawa-form
of the effective non-relativistic interaction for light degrees of freedom.}.

For example, if we assume a fifth force to be mediated through a novel spin-0 particle $X_0$ that couples to leptons and quarks with couplings $g_{l_i}$ and $g_{q_i}$ respectively, we can augment the Standard Model Lagrangian $\mathcal{L_\mathrm{SM}}$  to 
	\begin{equation}
	\mathcal{L} = \mathcal{L}_{\rm SM} +
	\sum_{i} \left[g_{l_{i}} \bar{l}_i   l_i + g_{q_{i}} \bar{q}_i   q_i \right ] X_0.
\label{eq:intnew}
\end{equation}
Here $i$ denotes the three flavor generations and $l_i$ and $q_i$ refer to the mass basis of the SM fermions. We note that the interactions of Eq.~(\ref{eq:intnew}) could be straightforwardly extended to (axial)vector or pseudoscalar particles and to flavor off-diagonal interactions, e.g. $g_{q_{ij}} \bar{q}_i q_j X_0$ with $i \neq j$. 
Further, we should emphasize that the operators of Eq.~(\ref{eq:intnew}) are gauge invariant only after electroweak symmetry breaking, which implies that the coefficients $g_{f_{ij}}$ must implicitly contain a factor $v/\Lambda_{\rm NP}$ ($v$ is the vacuum expectation value of the Higgs field and $\Lambda_{\rm NP}$ is a new physics scale). However, this is only important for the interpretation of the observed limit we derive on $g_{f_{i}}$. Studying Rydberg states in hydrogen atoms, we will set a limit on the combined interaction $g_e g_N$, where $g_e$ and $g_N$ corresponds to the interaction of $X_0$ respectively with the electron and the nucleon.

With the Lagrangian of Eq.~(\ref{eq:intnew}), the interaction mediated by the
NP boson $X_0$ between these two particles contributes an additional Yukawa potential $V(r)$ to the Hamiltonian.
Denoting by $r$ the distance between the electron and the nucleon and by $m_{X_0}$ the mass of the particle,
\begin{equation}
V(r) = (-1)^{s+1}\frac{g_e g_N}{4 \pi}~\frac{1}{r}~e^{-m_{X_0} r},
\label{eq:yukpot}
\end{equation}
where $s$, an integer, is the spin of the force mediator (e.g., $s=0$ for a scalar particle).
Higher integer-spin mediators would also give rise to a Yukawa potential of this form. There is however a subtle difference in the sign of this potential between even and odd integer-spin force carriers. Lorentz invariance and the unitarity of the transition matrix element lead to an attractive (repulsive) force if $g_e g_N > 0$ ($g_e g_N < 0$) in the case of an even-spin mediator, and to an attractive (repulsive) force if $g_e g_N < 0$ ($g_e g_N > 0$) in the case
of an odd-spin mediator. For example, as the charges for the Higgs boson (spin-0) and the graviton (spin-2) are the particles' masses, the Higgs force and gravity are both attractive. As we want to remain agnostic about the force carrier and the way it interacts with the nucleons and electrons, in the following we will allow both positive and negative values for $g_e g_N$.
Finally, we note that an excellent recent review of this type of simplified model is provided in \cite{Safronova2018}.

\section{New Physics level shifts}
\label{sec:NPshifts}

The presence of the interaction potential $V(r)$ would affect the atomic transition frequencies. Its effect can be evaluated perturbatively. To first order in $V(r)$, and neglecting spin-orbit coupling and other relativistic corrections, the energy
of a hydrogenic state of principal quantum number $n$, orbital angular momentum quantum number
$l$ and radial wave function $R_{nl}(r)$ is shifted by a quantity $\delta E_{nl}^{\rm NP}$, with
\begin{equation}
\delta E_{nl}^{\rm NP} = \int_0^\infty |R_{nl}(r)|^2 V(r)\, r^2\, dr.
\label{eq:NPshift}
\end{equation}
Since the interaction is spherically symmetric, the perturbation is diagonal in $l$ and in the magnetic quantum number $m$,
and $\delta E_{nl}^{\rm NP}$ does not depend on $m$.

Taking into account $V(r)$ to all orders, which we have done as a test of our numerical methods, confirms that second- and higher-order terms of the perturbation series are completely negligible for the couplings of interest, i.e. $|g_e g_N| < 10^{-11}$. 

The shift $\delta E_{nl}^{\rm NP}$ takes on a particularly simple form in the limit $m_{X_0}\rightarrow 0$: Since
\begin{equation}
|\delta E_{nl}^{\rm NP}| < \frac{|g_eg_N|}{4\pi} \int_0^\infty |R_{nl}(r)|^2\, \frac{1}{r}\, r^2\, dr,
\end{equation}
the virial theorem guarantees that 
\begin{equation}
|\delta E_{nl}^{\rm NP}| < \frac{|g_eg_N|}{4\pi}\,\frac{(-2E_n)}{\alpha Z},
\end{equation}
where $E_n$ is the non-relativistic energy of the
$(n,l)$ states, $\alpha$ is the fine structure constant and $Z$ is the number of protons in the nucleus. Moreover,
\begin{equation}
\lim_{m_{X_0}\rightarrow 0}
|\delta E_{nl}^{\rm NP}/E_n | = \frac{|g_eg_N|}{2\pi\alpha Z}.
\label{eq:smallmasses}
\end{equation}
(See Appendix~\ref{app:conversion} for
the origin of the factor of $1/\alpha$ and more generally for the conversion between natural units and atomic units.)
Simple analytical forms of $\delta E_{nl}^{\rm NP}$ can be obtained for states with maximum orbital angular momentum ($l=n-1$) or close to maximum orbital angular momentum --- see, e.g., Appendix~\ref{app:analyticalforms}. However,
in most cases $\delta E_{nl}^{\rm NP}$ is best evaluated numerically. 

Various approaches to this problem have been considered over the years, as has the calculation
of energy levels for a superposition of
a Coulomb potential and a Yukawa potential
(the Hellmann potential)
\cite{Adamowski85,Dutt86,Bag87,Hall01,Ikhdair07,Roy08,Nasser11,Ikhdair13,Onate16}. The most accurate results reported to date are those of Ref.~\cite{Roy08},
in which the energies of the ground state and first  few excited states
were obtained to approximately 13 significant figures using
a generalized pseudo-spectral method.
Our approach to this problem
is different and does not seem to have been used so far in this context:
We expand
the radial wave functions on a finite Laguerre basis of Sturmian functions
$S_{\nu l}^{\kappa}(r)$ \cite{Rotenberg62},
find the generalized eigenvectors of the matrix representing
the unperturbed Hamiltonian in that basis, and
use these to calculate the first order energy shift $\Delta E_{nl}$. Here
\begin{align}
S_{\nu l}^{\kappa}(r) = \sqrt{\frac{\kappa (\nu -1)!} {(\nu +l)(\nu +2l)!}}
(2\kappa r)^{l+1} &e^{-\kappa r} L_{\nu -1}^{2l+1}(2\kappa r), \nonumber \\
 & \nu=1,2,\ldots,
\end{align}
with $\kappa$ a positive parameter which can be chosen at will.
These 
basis functions have already been used in this context, but in a different way \cite{Nasser11}. Sturmian bases have proved to be convenient in precision
calculations of properties of hydrogenic systems \cite{Broad85,CPC}.

We obtain the eigenenergies and wave functions
of the unperturbed Hamiltonian by solving 
the generalized eigenvalue problem
\begin{equation}
{\sf H}_0{\sf c} = E\, {\sf Sc},
\label{eq:eigenvalueprob}
\end{equation}
where ${\sf H}_0$ is the matrix representing
the unperturbed non-relativistic Hamiltonian of hydrogen in this basis
and ${\sf S}$ is the overlap matrix of the basis functions (Sturmian
functions are not mutually orthogonal).
The corresponding matrix elements and the elements of the matrix ${\sf V}$ representing
the Yukawa potential
can be obtained in closed form using standard integrals
and recursion formula~\cite{Gradshteyn}. Having the eigenvectors ${\sf c}$,
the energy shifts are then calculated as $\delta E_{\sf c}^{\rm NP} = {\sf c^T V c}$.
Since the functions $\{S_{\nu l}^{\kappa}(r),
\nu=1,2,\ldots\}$ form a 
complete set, the eigenvalues $E$ and 
energy shifts $\delta E_{\sf c}^{\rm NP}$ obtained
with a basis of $N$ of these functions ($\nu =1,\ldots,N$)
converge variationally to the exact
eigenenergies and exact energy shifts when $N\rightarrow \infty$.
We repeat the calculations for several different values of $\kappa$ and
different basis sizes so as
to monitor the convergence of our results and the impact of
numerical inaccuracies. With an appropriate 
choice of $\kappa$, and taking $N$ up to 200, the calculated energy shifts converged
to at least 8 significant figures {}\footnote{See Supplemental Material at [URL, to be added by the Publisher] for tables of values of $\delta E_{nl}^{\rm NP}$ for
$n$ up to 80, $l$ up to 25, and
four different values of $M_{X_0}$ ranging from 1 to 1000~eV.}.
Using the same method, but solving the generalized eigenvalue problem for
the full Hamiltonian rather than the unperturbed Hamiltonian,
we could also reproduce 
the results of Ref.~\cite{Roy08} to the 14 significant figures given
in that article.

\begin{figure}[h]
\centering
\includegraphics[width=8.5cm]{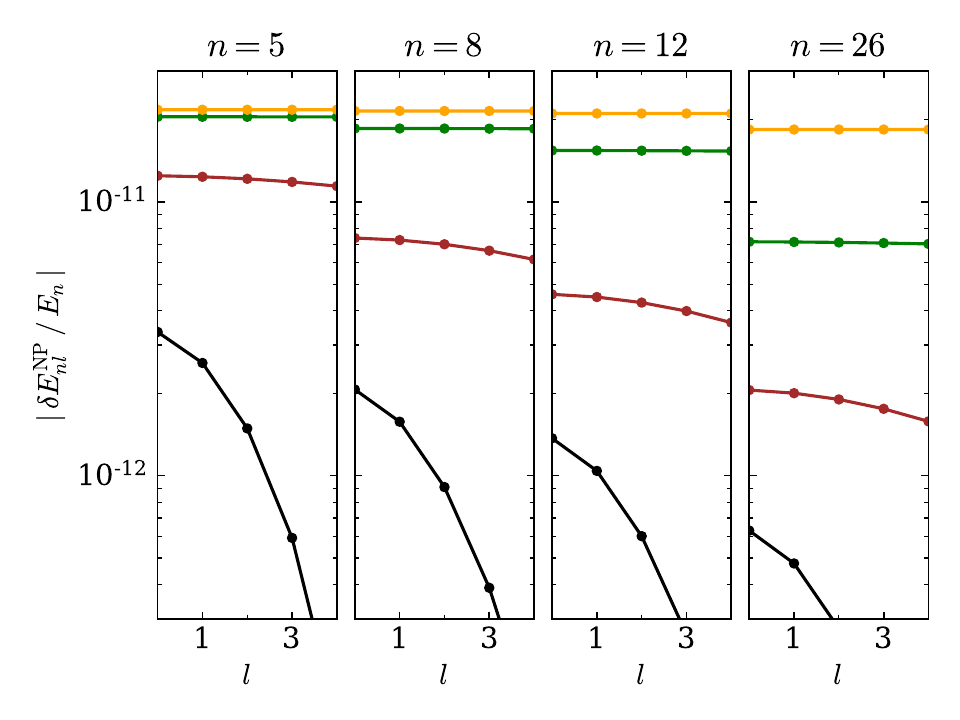}
\caption{The NP shift, $\delta E_{nl}^{\rm NP}$,
divided by the non-relativistic energy of the state, $E_n$,
for the states of atomic hydrogen with
$0 \leq l \leq 4$ and $n = 5$ ($E_n/h = -1.32\times10^{11}$~kHz, where $h$ is Planck's constant),
8 ($E_n/h = -5.14\times10^{10}$~kHz), 12 ($E_n/h = -2.28\times10^{10}$~kHz)
or 26 ($E_n/h = -4.87\times10^9$~kHz).
A value of $g_e g_N$ of $1 \times 10^{-12}$ is
assumed. From top to bottom, $m_{X_0} = 1$~eV (orange circles),
10~eV (green circles), 100~eV (brown circles), or 1~keV (black circles).
}
\label{fig:fig1}
\end{figure}

\begin{figure}[h]
\centering
\includegraphics[width=8.5cm]{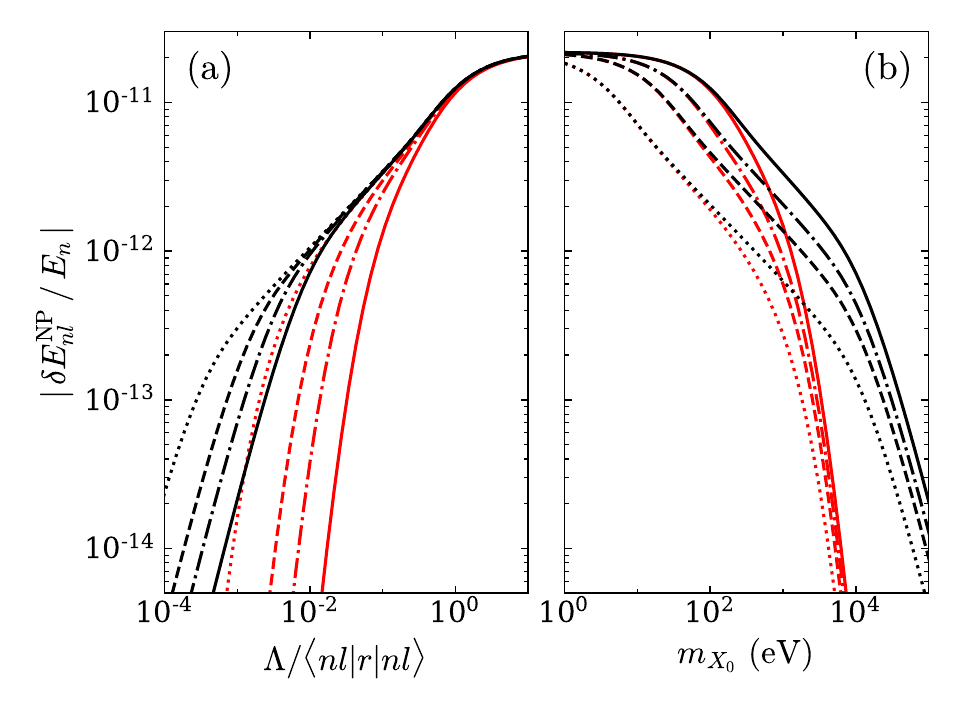}
\caption{The NP shift, $\delta E_{nl}^{\rm NP}$,
divided by the non-relativistic energy of the state, $E_n$,
for the states of atomic hydrogen with $n=5$ (solid curves),
$n=8$ (dashed-dotted curves), $n=12$ (dashed curves) or $n=26$
(dotted curves) and $l=0$ (black curves) or $l=2$ (red curves),
vs., (a) the range of the NP potential divided by the characteristic
length scale of the atomic wave function, 
$\langle nl | r | nl\rangle$, or
(b) the mass of the NP particle.
A value of $g_e g_N$ of $1 \times 10^{-12}$ is
assumed. 
}
\label{fig:varyingmass_581226}
\end{figure}

\begin{figure}[h]
\centering
\includegraphics[width=8.5cm]{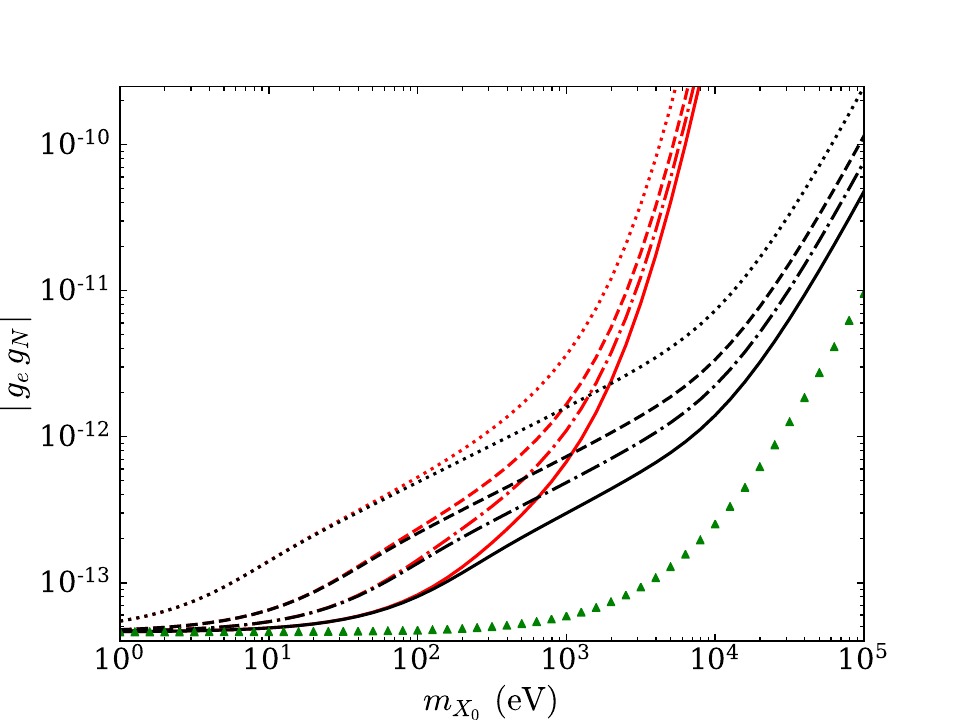}
\caption{The coupling constant $g_e\,g_N$ at which the
relative shift $|\delta E_{nl}^{\rm NP}/E_n|$ is $1\times 10^{-12}$, vs.\ 
the mass of the NP particle. The line styles and colors are the same as 
in Fig.~\ref{fig:varyingmass_581226}.
Green triangles: the results for the ground state.
}
\label{fig:fixedrelshift_581226}
\end{figure}

The results of these calculations are summarized in 
Figs.~\ref{fig:fig1}, \ref{fig:varyingmass_581226} and \ref{fig:fixedrelshift_581226}. These results, like
all the other numerical results discussed in this paper, refer to the specific case of atomic hydrogen. We will therefore assume that $Z=1$ from now on.

Fig.~\ref{fig:fig1} shows the general trends. The fractional shift is largest for light bosons, where the range of the Yukawa potential is comparable to or larger than the range of the atomic wave function. In agreement with Eq.~(\ref{eq:smallmasses}),
$|\delta E_{nl}^{\rm NP}/E_n| \lesssim |g_eg_N|/2\pi\alpha$ for low masses.
As $m_{X_0}$ increases, the shift decreases, but in a way that depends on the shape of the atomic wave function through both $n$ and $l$. The effect of the Yukawa potential is largest at the origin. As $n$ and $l$ increase, the probability density of the atomic wave function in the region close to the nucleus is reduced, leading to a smaller NP shift.

\renewcommand{\arraystretch}{1.2}
\begin{table}
\caption{The range of the Yukawa potential ($\Lambda$), expressed as a multiple of the Bohr radius, and the principal quantum number $n_\Lambda$
for which this range is equal to that of the corresponding $l=0$ state to the closest
approximation possible, for three values of
$m_{X_0}$, the mass of the NP particle.}

\label{table:C}
\begin{center}
\begin{ruledtabular}
\begin{tabular}{p{2cm}p{3cm}p{0.7cm}}
$m_{X_0}$ & $\Lambda$ & $n_\Lambda$\\
\tableline\\[-4mm]
1 eV & $3.73 \times 10^{3}\, a_0$ &
50 \\
100 eV & $3.73 \times 10^{1}\, a_0$ & 5 \\
10 keV & $3.73 \times 10^{-1}\, a_0$ & 1 
\end{tabular}
\end{ruledtabular}
\end{center}
\end{table}
To gain further insight, in Fig.~\ref{fig:varyingmass_581226}, we investigate the relationship between the two characteristic length scales of the problem, i.e., the range of the Yukawa potential, $\Lambda = 1/m_{X_0}$, and the range of the atomic wave function.
The latter can be characterized by the expectation value $\langle nl | r | nl \rangle$, which for $l=0$ states is $3a_0n^2/2$ where
$a_0$ is the Bohr radius. We see that these two ranges are comparable for principal quantum numbers $n \sim n_\Lambda$, where $n_\Lambda$ is the integer closest to $(2\,\Lambda/3\,a_0)^{1/2}$. Representative values of $n_\Lambda$ are given in
Table~\ref{table:C}. The NP shift is accurately
predicted by Eq.~(\ref{eq:smallmasses}) for $n\ll n_\Lambda$ and is much smaller than that limit
for $n\gg n_\Lambda$. 
The fractional shift is plotted in Figs.~\ref{fig:varyingmass_581226}(a) and (b), respectively
against the ratio of these two characteristic lengths 
and against the boson mass, for a range of values of $n$ and $l$.
These curves show that for masses below $\sim 50$~eV, the shift decreases with $n$ but is essentially independent of $l$. Above this breakpoint, the shift decreases much more rapidly for d-states ($l=2$) than for s-states ($l=0$). This trend is even more marked for higher values
of $l$ (not shown in the figure). In fact, for states with $l = n-1$ (which is the maximum value of
the orbital angular momentum for the principal quantum number $n$),
$|\delta E_{nl}^{\rm NP}/E_n|$ decreases as fast as $n^{-2n}$ when $n$ increases beyond $n_\Lambda$
(see Appendix~\ref{app:analyticalforms}).

In Figure~\ref{fig:fixedrelshift_581226}, we fix the value of the fractional NP shift at $|\delta E_{nl}^{\rm NP}/E_n| = 10^{-12}$, and show how the resulting constraint on the mass $m_{X_0}$ and the effective coupling $g_e g_N$ depend on the quantum numbers $n$ and $l$. Thus combining measurements for different values of $n$ and $l$ could provide additional information on the properties of the fifth-force carrier, i.e., its mass and its couplings to the electron and nuclei.

\section{NP bounds based on current spectroscopic data}
\label{section:current}

In a nutshell, the existence of a new physics interaction could be brought to light by demonstrating a significant difference between the measured transition frequency for a transition from a state $a$ to a state $b$, $\Delta_{ba}^{\rm exp}$, and the corresponding prediction of the Standard Model, $\Delta_{ba}^{\rm SM}$ (or, better, by demonstrating such a difference for a set of transitions). Bounds on the strength of the new physics interaction can be set by finding the most positive and most negative values of $g_eg_N$ for which $\Delta_{ba}^{\rm exp}$ is consistent with the theoretical value $\Delta_{ba}^{\rm SM} + \Delta_{ba}^{\rm NP}$ with
\begin{equation}
    \Delta_{ba}^{\rm NP} = (\delta E_{n_bl_b}^{\rm NP} - \delta E_{n_al_a}^{\rm NP})/h,
\end{equation}
where $h$ is Planck's constant.
However, $\Delta_{ba}^{\rm SM}$ depends on the Rydberg constant, $R_{\infty}$, and usually also on the charge radius of the nucleus, whose values are primarily obtained by matching spectroscopic data to theory \cite{Codata2010,Codata2014}.  Setting bounds on $g_eg_N$ makes it therefore necessary to evaluate $\Delta_{ba}^{\rm SM}$ with these constants set to values themselves obtained with allowance made for the possibility of new physics shifts on the relevant atomic transitions.
Frequency intervals have been both measured and calculated to a very high level of precision for transitions in hydrogen, deuterium and muonic hydrogen. However, a new physics interaction might couple an electron differently to a deuteron than to a proton, and couple a proton differently to a muon than to an electron. It is therefore prudent, when establishing such bounds, to use data pertaining to only one of these three systems rather than using mixed sets of data. We consider bounds based exclusively on hydrogen results in this paper.

$\Delta_{ba}^{\rm SM}$ is the sum of a gross structure contribution 
$\Delta_{ba}^{\rm g}$ (as given
by the elementary treatment based on the Schr\"odinger equation)
and of various corrections arising from the Dirac equation, from QED effects
and from the hyperfine coupling \cite{Eides:2000xc,Karshenboimreview,Codata2010,Codata2014,Yerokhintables,Horbatsch2016,Yerokhin2019}.
In terms of the Rydberg frequency, ${\cal R} = c\,R_{\infty}$,
\begin{equation}
    \Delta_{ba}^{\rm g} = {\cal R}\,\left(\frac{1}{n_a^2}-\frac{1}{n_b^2}\right)\,
    \frac{m_{\rm r}}{m_e},
\end{equation}
where $m_{\rm r}$ is the reduced mass of the atom and $m_e$ is the mass of the electron. It 
is convenient to factorize $\Delta_{ba}^{\rm g}$ into the product
${\cal R}\,\tilde{\Delta}_{ba}^{\rm g}$, with 
\begin{equation}
    \tilde{\Delta}_{ba}^{\rm g} = \left(\frac{1}{n_a^2}-\frac{1}{n_b^2}\right)\,
    \frac{m_{\rm r}}{m_e}.
\end{equation}
The difference $\Delta_{ba}^{\rm SM} - \Delta_{ba}^{\rm g}$ depends on $R_p$, the charge radius
of the proton, through a term roughly proportional to $R_p^2$
\cite{Eides:2000xc,Karshenboimreview,Codata2010,Codata2014,Yerokhintables,Horbatsch2016,Yerokhin2019}. We denote this term by
$R_p^2\,\tilde{\Delta}_{ba}^{\rm ns}$, aggregate all the other corrections into
a shift $\Delta_{ba}^{\rm oc}$, and
write
\begin{equation}
    \Delta_{ba}^{\rm SM} = {\cal R}\,\tilde{\Delta}_{ba}^{\rm g}
    + R_p^2\,\tilde{\Delta}_{ba}^{\rm ns} + \Delta_{ba}^{\rm oc}.
    \label{eq:Deltath}
\end{equation}
The term $\Delta_{ba}^{\rm oc}$ includes fine structure and recoil corrections as well as QED and hyperfine shifts. Detailed work by a number of authors has yielded expressions for these corrections in terms of ${\cal R}$, of $R_p$ and of a small number of fundamental constants determined from measurements in physical systems other than hydrogen. 
The values of ${\cal R}$ and $R_p$ recommended by the Committee on Data of the International Council for Science (CODATA) were co-determined by a global fit of the theory to a large set of data, including deuterium data \cite{Codata2014}. Taking new physics shifts into account in a determination of ${\cal R}$ based entirely on hydrogen data thus involves a simultaneous redetermination of $R_p$.
Eq.~(\ref{eq:Deltath}) is a convenient starting point for such calculations {}\footnote{
The terms $R_p^2\,\tilde{\Delta}_{ba}^{\rm ns}$ and $\Delta_{ba}^{\rm oc}$ are also proportional to
${\cal R}$; however, their dependence in ${\cal R}$ is normally not important
for the determination of this constant as these terms are much smaller than ${\cal R}\,\tilde{\Delta}_{ba}^{\rm g}$ unless $n_a = n_b$. $R_p^2\,\tilde{\Delta}_{ba}^{\rm ns}$ is effectively
zero for $l\not= 0$. For $l = 0$, this term depends on the proton radius both through the overall
factor $R_p^2$ and through a dependence of $\tilde{\Delta}_{ba}^{\rm ns}$ on $R_p$; however,
the latter dependence is weak and does not complicate the calculation.}.

Bearing this in mind, we derive bounds on the value of $g_eg_N$ in the
following way: Given experimental transition
frequencies for several different intervals, e.g., 
$\Delta_{b_1a_1}^{\rm exp}$, 
$\Delta_{b_2a_2}^{\rm exp}$, 
$\Delta_{b_3a_3}^{\rm exp}$, etc.,
we calculate a value of ${\cal R}$ and a value of $R_p$
by matching these results with the corresponding theoretical frequency intervals,
\begin{align}
\Delta_{b_ia_i}^{\rm th} = 
    {\cal R}\,\tilde{\Delta}_{b_ia_i}^{\rm g}
    + R_p^2\,\tilde{\Delta}_{b_ia_i}^{\rm ns} + & \Delta_{b_ia_i}^{\rm oc} + \Delta_{b_ia_i}^{\rm NP}, \nonumber\\
    & \; \qquad i = 1,2,3,\ldots
    \label{eq:ansatz}
\end{align}
The values of these two parameters are determined by correlated $\chi^2$-fitting.
We then obtain bounds on the coupling constant 
by finding the most positive and most negative values of
$g_eg_N$ for which the model fits the data at the 5\% confidence level. The sensitivity to new physics arises because of the dependence of the NP shift on the quantum numbers $n$ and $l$ illustrated in Figs. 1-3. Put simply, states with high values of $n$ and $l$ are only weakly sensitive to new physics, whereas the opposite is the case for low-lying states. 

Before describing the results of this analysis, we briefly discuss the existing experimental results relevant for this calculation and the related theoretical uncertainties. Further details about the calculation can be found in Appendix \ref{app:fittingcurr}.

\subsection{Existing spectroscopic data for hydrogen}

Clearly, detecting a NP interaction from spectroscopic data sets a challenging
level of precision and accuracy on the measurements. Apart from the hyperfine splittings of the 1s and 2s states, which are not directly relevant here,
the only hydrogen frequency intervals currently known to an accuracy better than 1~kHz are the 
1s~--~2s
interval, which has been measured with an experimental error of 10~Hz (i.e., a relative error of 0.004 ppt) \cite{Parthey2011,Matveev2013},
and intervals between circular states with $n$ ranging from 27 to 30, for which unpublished measurements with an experimental error
of a few Hz (about 10~ppt) have been made \cite{DeVries2002}. Circular states are states with $|m| = l = n-1$.

The recommended value for the Rydberg constant
is based on the 1s~--~2s measurement as well as on a number of measurements with a larger error \cite{Codata2014,Codata2017}.
The latter include measurements of the 
2s~--~8s, 2s~--~8d and 2s~--~12d intervals
made in the late 1990s with an experimental error ranging from 6 to 9~kHz (i.e., of the order of 10~ppt)
\cite{deBeauvoir1997,Schwob1999, deBeauvoir2000}.
Until recently, no other transitions between hydrogen states differing in $n$ had been measured with an error of less than 10~kHz. However, the centroid of the 2s~--~4p
interval has now been determined
with an error of 2.3~kHz \cite{Beyer2017}, and that of the 1s~--~3s interval with an error of 2.6~kHz
(1~ppt) \cite{Fleurbaey:2018fih}.

\subsection{Theoretical uncertainty}

The overall uncertainty on the SM predictions of hydrogen energy levels is mainly contributed by uncertainties on the values of the Rydberg constant, of the proton radius and of various QED corrections. Uncertainties on the
values of other fundamental constants also contribute, although not in a significant way at the level of precision these energy levels can currently be calculated. 

The uncertainty on the values of
the Rydberg constant and the proton radius does not affect our calculation of the
bounds on $g_eg_N$ (recall that within our approach, these values are determined together with the bounds themselves in a self-consistent way).

As is well known, the SM theory of the energy levels of hydrogen has developed enormously since the early days of Quantum Mechanics \cite{Eides:2000xc,Karshenboimreview}. Compilations of the relevant QED and hyperfine corrections and their uncertainties have been published, e.g., by the CODATA collaboration \cite{Codata2010,Codata2014}, and recent updates 
can be found in \cite{Yerokhintables}, \cite{Horbatsch2016} and \cite{Yerokhin2019}.
These corrections roughly scale as $n^{-3}$ and strongly depend on $l$. Ref.~\cite{Horbatsch2016} gives the
combined theoretical uncertainty on the energy of a state of principal quantum number $n$ as $(2.3/n^3)$~kHz for $l=0$, excluding the error contributed by the uncertainty on $R_p$, and as less than 0.1~kHz for $l > 0$. More recent work has lowered this uncertainty. For example,
Ref.~\cite{Yerokhin2019} gives it as $(1.8/n^3)$~kHz for $l=0$,
excluding the error arising from the uncertainty on $R_p$, and
further progress in this direction has been made since (e.g.,
\cite{Karshenboim2019a,Karshenboim2019b,Karshenboim2019c,Karshenboim2019d}).
Except for the 1s~--~2s interval, the experimental uncertainty rather than the theoretical uncertainty is thus the main limitation for setting bounds on $g_eg_N$ based on the current spectroscopic data.

\renewcommand{\arraystretch}{1.2}
\begin{table}[t]
     \caption{Values of the Rydberg frequency 
     obtained by previous authors or derived in this work, assuming no NP interaction. The numbers between parentheses are the uncertainties on the last digit quoted.}
    \label{table:R}
    \begin{center}
    \begin{ruledtabular}
    \begin{tabular}{ll}
      Reference & 
      ${\cal R}
      $\\ \tableline\\[-4mm] 
      \multirow{1}{*}{CODATA 2014 \cite{Codata2014}} 
      & $\mbox{3 289 841 960 355(19)}~\mbox{kHz}$ \\[0mm]
       \multirow{1}{*}{Beyer {\it et al.} \cite{Beyer2017}} 
      & $\mbox{3 289 841 960 226(29)}~\mbox{kHz}$ \\[0mm]
        \multirow{1}{*}{Fleurbaey {\it et al.} \cite{Fleurbaey:2018fih}} 
        & $\mbox{3 289 841 960 362(41)}~\mbox{kHz}$ \\[0mm]
        \multirow{1}{*}{De Vries \cite{DeVries2002}} 
        & $\mbox{3 289 841 960 306(69)}~\mbox{kHz}$ \\[0mm] 
\multirow{1}{*}{Dataset A} 
        & $\mbox{3 289 841 960 306(18)}~\mbox{kHz}$\\[0mm]
\multirow{1}{*}{Dataset B} 
        & $\mbox{3 289 841 960 356(23)}~\mbox{kHz}$\\[0mm]
        \multirow{1}{*}{Dataset C} 
        & $\mbox{3 289 841 960 307(18)}~\mbox{kHz}$
    \end{tabular}
    \end{ruledtabular}
\end{center}
\end{table}
\subsection{Bounds based on existing data}
\begin{figure}[t]
\centering
\includegraphics[width=8.5cm]{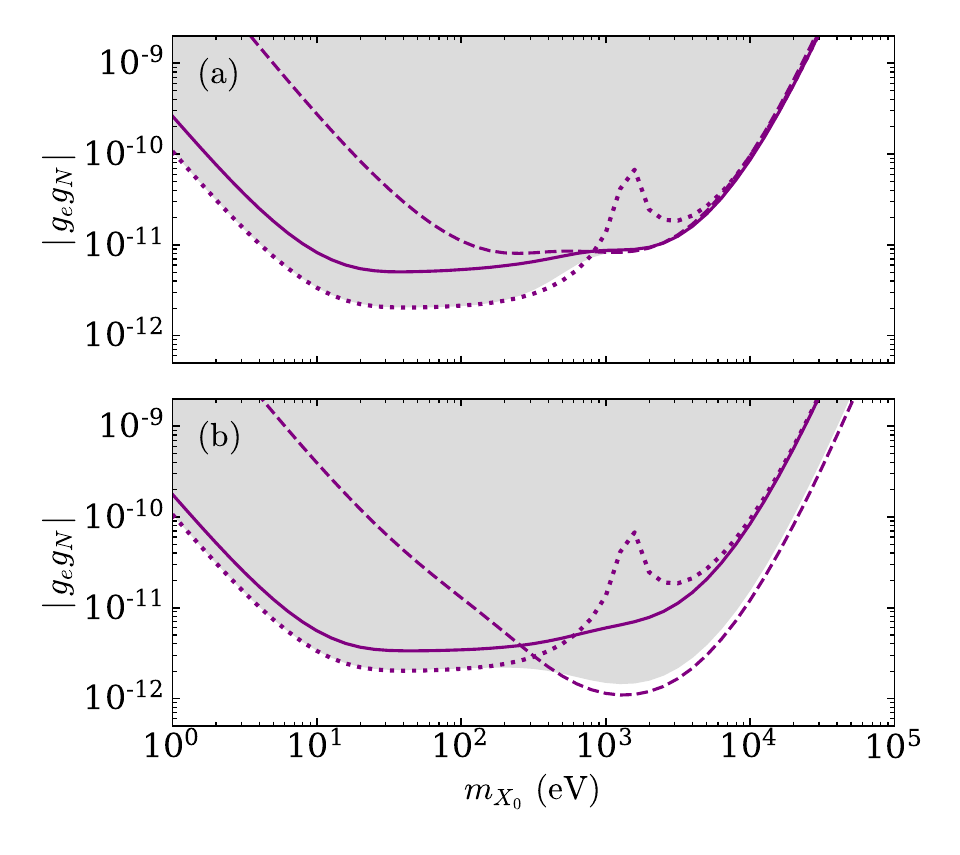}
\caption{Upper bounds on the possible value of $|g_e g_N|$ derived
from existing spectroscopic data, (a) for
an attractive interaction, (b) for a repulsive interaction. Shaded areas: region excluded at the 95\% confidence level (data set A).
Solid and dashed curves: bounds based on the same set of transitions as for the shaded areas, minus the 2s~--~4p transition (data set B, solid curves) or minus the transitions between high lying circular states (data set C, dashed curves). Dotted curves: bounds arising from a comparison of
experimental results for these high lying circular states to theoretical predictions based on the data set C.
}
\label{fig:bounds_paper}
\end{figure}
Bounds on the NP interaction strength derived as explained above are presented in
Figs.~\ref{fig:bounds_paper}(a) and \ref{fig:bounds_paper}(b), respectively for attractive and repulsive interactions. These results are based on three different sets of data, which we refer to as sets A, B and C. Set A groups all the existing high precision spectroscopic measurements in hydrogen, namely all the 18 experimental hydrogen transition frequencies included in the CODATA~2014 least square fit~\cite{Codata2014},
the recent results of Ref.~\cite{Beyer2017} for the 2s~--~4p interval and of Ref.~\cite{Fleurbaey:2018fih} for the 1s~--~3s interval, and
the results of Ref.~\cite{DeVries2002} for the transitions between high lying circular
states. The other two sets are the same as Set A but without the 2s~--~4p results (Set B) or without the circular states results (Set C). The corresponding values of ${\cal R}$ obtained when assuming no NP shift are 
given in Table~\ref{table:R}, together with the recommended value of this constant \cite{Codata2014}, values based on the recent measurements of either the 2s~--~4p or the 1s~--~3s intervals \cite{Beyer2017,Fleurbaey:2018fih} and a value based
entirely on measurements of transitions between the circular states \cite{DeVries2002}.  As is well known, the results of Ref.~\cite{Beyer2017} are discrepant with both the CODATA results and those of Ref.~\cite{Fleurbaey:2018fih} in regards to the values of ${\cal R}$ and $R_p$, but yield a value of $R_p$ in good agreement with measurements in muonic hydrogen \cite{Antognini2013}. 
The values of ${\cal R}$ obtained from Dataset B are in close agreement with the CODATA 2014 value and have an uncertainty of a similar magnitude, although the CODATA fit also included spectroscopic measurements in deuterium and scattering data. Including the results of Ref.~\cite{Beyer2017} in the fit reduces 
${\cal R}$ significantly (the change is large because of the particularly small experimental error on these measurements). 

Our main results for the current bounds on $g_eg_N$ are based on Dataset A and are represented by the shaded
areas in Fig.~\ref{fig:bounds_paper}. They set a constraint of better than $10^{-11}$ over the range of $10^1$~--~$10^3$~eV. As seen from the figure, the shape of the excluded area somewhat differs between attractive and repulsive interactions, particularly in the region around 100 eV. This difference indicates that the range of allowed values of  $g_eg_N$ is not centred on zero --- though we emphasise that a value of zero remains compatible with the experimental data. The regions below the shaded areas indicate the range of values of $g_eg_N$ compatible with the data, given the experimental and theoretical errors {}\footnote{These results are consistent with the hydrogen bound proposed in Ref.~\cite{Karshenboim:2010cg}, which was obtained in a different way.}.

Next we consider the effect of removing individual measurements from the calculation. Removing the recent 2s~--~4p measurement \cite{Beyer2017} has a considerable effect, not only weakening the overall bound, as expected, but also changing the shape of the excluded region. These differences reflect the aforementioned inconsistencies in the values of the Rydberg constant and the proton radius derived from the results of Ref.~\cite{Beyer2017} with those obtained in the CODATA 2014 fit. The effect of removing this measurement illustrates the perils of selectively setting bounds using individual measurements or combinations of measurements. Whilst individual measurements may be precise, their accuracy can only be gauged against other measurements, particularly independent measurements of the same transitions.

Instead of removing the 2s~--~4p measurements, we now remove the unpublished circular state measurements of  Ref.~\cite{DeVries2002} and use Dataset C. The result is a substantial weakening of the NP bound for lower masses,  illustrating the importance of using
measurements of
states with a large spatial extension when probing for a NP interaction with a low value of $m_{X_0}$ \cite{Karshenboim:2010cg}. Although small, the NP shift of the circular states is not negligible when the range of
the interaction is long enough. This leads to a decrease in the relative shift of these states compared to the
low lying states when $m_{X_0}\rightarrow 0$, and hence to a weakening of the bounds on $|g_eg_N|$ {}\footnote{Since completion of this work we have become aware of a new  measurement of the 2s$_{1/2}$~--~2p$_{1/2}$ Lamb shift [N.~Bezginov, T.~Valdez, M.~Horbatsch, A.~Marsman, A.~C.~Vutha and E.~A.~Hessels, {\it A Measurement of the Atomic Hydrogen Lamb Shift and the Proton Charge Radius}, Science {\bf 365}, 1007 (2019)]. Adding this result to those already included in Dataset A and carrying out the global fitting procedure on this extended set yields bounds incompatible with a zero value of $g_eg_N$ at the 95\% confidence level. It is our opinion that this surprising result merely reflects the inconsistencies in the data noted in the text, which are exacerbated by the inclusion of this new measurement. 
}.

In summary, we have derived global NP bounds based on all available measurements for hydrogen, with no input from other atomic species. The sensitivity of the bound to individual measurements and to the Rydberg constant illustrates that bounds set using measurements on individual transitions should be treated with a degree of caution. The strong additional constraint provided by high-lying states at low masses motivate precision measurements for states with both higher $n$ and $l$. For the latter we note the proposal of the Michigan group \cite{Ramos2017}.

Before closing this section, we note that bounds on $g_eg_N$ can also be found by
comparing the values of ${\cal R}$ derived from different sets of transitions. 
For example, let ${\cal R}_C(m_{X_0},g_eg_N)$ and ${\cal R}_{D}(m_{X_0},g_eg_N)$ be the NP-dependent values of ${\cal R}$ obtained by fitting the theoretical
model respectively to Dataset C and to the circular state results of Ref.~\cite{DeVries2002}. These two sets of data are completely independent of each other, and by contrast to
${\cal R}_C(m_{X_0},g_eg_N)$,
the calculation of ${\cal R}_{D}(m_{X_0},g_eg_N)$ is insensitive to uncertainties on the proton radius and to poorly known QED corrections.  The corresponding errors on these Rydberg frequencies,
$\sigma_C$ and $\sigma_{D}$, are also functions of
$m_{X_0}$ and $g_eg_N$. As these errors are not correlated
with each other, bounds on the NP coupling constant
can be obtained by finding the most positive and most negative values of
$g_eg_N$ such that
\begin{equation}
|{\cal R}_C(m_{X_0},g_eg_N) - 
{\cal R}_{D}(m_{X_0},g_eg_N)|
 = f\,\sqrt{\sigma_C^2 +\sigma_{D}^2}
\label{eq:bounds2}
\end{equation}
for a given choice of $f$ (this constant sets the confidence limit of the bounds --- we take
$f=2$). The results are also shown in Fig.~\ref{fig:bounds_paper} (the dotted curves). Cancellations of NP shifts are at the origin of the large weakening of these bounds between 1 and 10~keV. They are similar, below 300~eV, to those obtained from the global fit of the same set of data (the shaded areas). Compared to a global fit, however, this approach to setting bounds is potentially more sensitive to systematic errors in some of the measurements. We thus prefer to take the shaded areas as the best representation of the constraint on $g_eg_N$ that can be set on the basis of the current body of spectroscopic work in hydrogen.

\section{Scope for tighter bounds}
\label{sec:improve}
Three factors limit the strength of the current bound shown in Fig.~\ref{fig:bounds_paper}. The first is the experimental uncertainty of the measured energy levels. So far, only the 1s~--~2s interval has been measured with a relative uncertainty below the 0.01~ppt level. For higher states such as the measurements at $n=12$, the $\sim 1$ kHz uncertainty is approximately one hundred times larger or more. The second factor is the range of quantum numbers $n$ and $l$ for which precise data exist. The importance of additional measurements is highlighted in Fig.~\ref{fig:bounds_paper}. Lastly, the limitations on the SM calculation of the energies also plays an important role. Here also there is much to be gained by working with higher-lying Rydberg states. In this section we consider the prospects for improvements in each of these three areas.

\subsection{Improved measurements}
In this section we consider the effect of reducing the current experimental uncertainty approximately 100-fold, such that all transition frequencies in the dataset are known to the 10~Hz level currently available for the 1s~--~2s interval.  As an aspirational goal we also consider what could be achieved with measurements at the 1~Hz level. A detailed discussion of future experiments is outside the scope of this article. Here we briefly discuss the dominant sources of  uncertainties with the 10~Hz goal in mind. The focus is on laser spectroscopy of low-$l$ states; improved measurements of circular Rydberg states are considered in \cite{Ramos2017}.

The current measurement uncertainty includes contributions from both the background electromagnetic environment and atomic motion. Fundamental limits are provided by the radiative linewidth and black-body radiation (BBR). We calculated the radiative width and  black-body shift and broadening of the relevant states (Appendix \ref{BBR}).
At $n=9$, the radiative linewidth (which varies as $n^{-3}$) is approximately~100 kHz for the s state and roughly ten times larger for the d state. The simple lineshape when radiative broadening dominates should enable line centres to be determined with high accuracy, with recent measurements in hydrogen determining line centres to one part in 10,000 of the linewidth \cite{Beyer2017}. As described in Appendix \ref{BBR}, we find that black-body related uncertainties can be neglected even at 300~K provided that the temperature can be stabilised to 0.01~K.

Concerning stray fields, we note that the magnetic moment of low $l$ states does not vary with $n$. Therefore, methods developed for precision measurements with low $n$ states can be applied. For s-states the very small differential Zeeman shift is easily controlled at the sub-Hz level \cite{Huber1999,deBeauvoir2000}, while for d-states differential measurements such as those routinely carried out in optical atomic clocks \cite{Derevianko2011} can be used to largely eliminate magnetic field errors. A much greater challenge is presented by the DC Stark shift, which scales as $n^2$ and $n^7$ for the linear and quadratic components respectively. A detailed analysis of the effect of the DC Stark shift on the hydrogen Rydberg spectrum is provided in \cite{deBeauvoir2000}. In their experiments a stray field of $\sim$3~mV~cm$^{-1}$ was reported, leading to a final contribution to the uncertainty at the kHz level. However other experiment have shown that stray fields can be reduced to the 30~$\upmu$V~cm$^{-1}$ level by performing electrometry with high-$n$ states ($n>100$) \cite{Frey1993,Osterwalder1999}. Drift rates as low as 2~$\upmu$V~cm$^{-1} \mathrm{h}^{-1}$ have also been measured \cite{Hogan2018}. Such measurements could be performed independently using a co-electrometry with a different species \cite{Osterwalder1999,Hogan2018}. For a field of 30~$\upmu$V~cm$^{-1}$, the quadratic Stark effect is dominant for s-states, and measurements with 10 Hz uncertainty should be possible up to $n=23$, with $n\approx 40$ accessible if the stray field is determined to 1~$\upmu$V~cm$^{-1}$. For d-states, the linear Stark effect dominates, but differential measurements between different $|m|$ states should enable the first order shift to be cancelled. The resulting uncertainty thus becomes dominated by the residual quadratic shift. 

Considering motional effects, we note that all measurements of hydrogen energy levels to date have been  performed in atomic beams, where second-order Doppler effects limit the achievable linewidth to approximately 1~MHz. A complex velocity-dependent lineshape analysis is thus required to extract the true line center to the current 1~kHz accuracy \cite{deBeauvoir2000}. 

In other atomic species, using ultracold atoms has enabled a dramatic reduction in the uncertainty of optical frequency measurements. Sub-10 Hz uncertainty has been achieved with untrapped atoms \cite{Wilpers2007}, while measurements based on atoms confined in magic-wavelength traps are entirely limited by the uncertainty in the microwave-based definition of the SI second \cite{campbell2008,Kim2017}.

For Rydberg states, experiments with ultracold atoms are dominated by the large level shifts due to the long-range van der Waals interaction \cite{Beguin2013}, which scales as $n^{11}$. Control over the number of atoms and interparticle distance and geometry is therefore essential.  Confining atoms to a volume of $\sim 1\  \upmu \mathrm{m}^3$ would also largely eliminate errors due to field gradients. Therefore, a suitable platform could consist of individual hydrogen atoms confined in a single optical tweezer or tweezer array. Single-atom arrays have now been achieved with a growing range of atomic \cite{Bergamini2004,Cooper2018,Norcia2018,Saskin2019} and even molecular \cite{Liu2018,Anderegg2019} species.  Substantial hurdles exist for realising a similar system in hydrogen, not least the difficulty of laser cooling \cite{Setija1993}, which has so far proven essential for loading the optical tweezers. However alternative approaches such as loading from a hydrogen Bose-Einstein condensate \cite{Fried1998}, careful dissociation of laser-cooled hydride molecules \cite{Lane2015} or in-trap Sisyphus cooling \cite{Saijun2011} may also provide possible routes. Here we assume that such a system may be realised, and that the contribution of the Doppler and recoil effects can be reduced below the natural linewidth of the transition by using well established two-photon spectroscopy techniques \cite{deBeauvoir2000}, possibly in combination with resolved sideband cooling \cite{Kaufman2012}. Trap-induced AC Stark shifts are eliminated by extinguishing the trap light during the spectroscopy, as is common in Rydberg experiments with tweezer arrays.

Overall, we consider that a target of extending the range of states measured with an absolute uncertainty of 10 Hz or better to the full Rydberg series of s- and d-states up to a principal quantum number of $n\approx 40$ is feasible.  We note that this is still some way off the spectroscopic state-of-the-art achieved with cold trapped atoms. For circular states, 10~Hz uncertainty has already been achieved \cite{DeVries2002}; here achieving a precision of 0.1 Hz in future measurements seems feasible.

\subsection{Improved theory}
\label{section:improvedtheory}
Improved measurements at the 10~Hz level would also provide a challenge to the current theory of SM corrections to hydrogen energy levels. Uncertainties in $\cal{R}$ and $R_p$ could be removed by using the global fitting procedure described in Section \ref{section:current}. Concerning the remaining correction due to QED and other effects, we note that the current uncertainty on the Lamb shift of the 2p$_{1/2}$ state is 21~Hz, including the uncertainty on the shift of the centroid of that level due to the hyperfine coupling \cite{Yerokhin2019}. As the theoretical error on QED and hyperfine corrections scales roughly like $1/n^3$ and has been found to be smaller for states with larger orbital angular momentum, the theoretical error for the states with $l > 0$ is already expected
to be below 10~Hz for $n \geq 3$ and below 1~Hz for $n \geq 6$. The situation for s-states is less clear. Current work assumes that the error on these corrections scales as $n^{-3}$, at least down to the 100~Hz level  \cite{Horbatsch2016,Yerokhin2019}. Given the current theoretical uncertainty on the energy of the 2s state, achieving an accuracy of 10~Hz would require a considerable effort in the evaluation of QED corrections that are currently poorly known. Alternatively, the data may be fitted to a theoretical model which does not rely on values of the Lamb shift accurate to the 10~Hz level but instead treats the theoretical error on this quantity as a fitting parameter, assuming a $n^{-3}$ scaling beyond the corrections that could be calculated. We used such a model to obtain the illustrative results presented in Section~\ref{NumIll} (the method is outlined in Appendix~\ref{app:fittingproj}).
However, further theoretical work would be necessary
to confirm that the assumed $n^{-3}$ scaling still holds, in sufficiently good approximation, down to errors as small as 10~Hz or less.

\subsection{Numerical illustration}
\label{NumIll}

\begin{figure}[t]
\centering
\includegraphics[width=8.5cm]{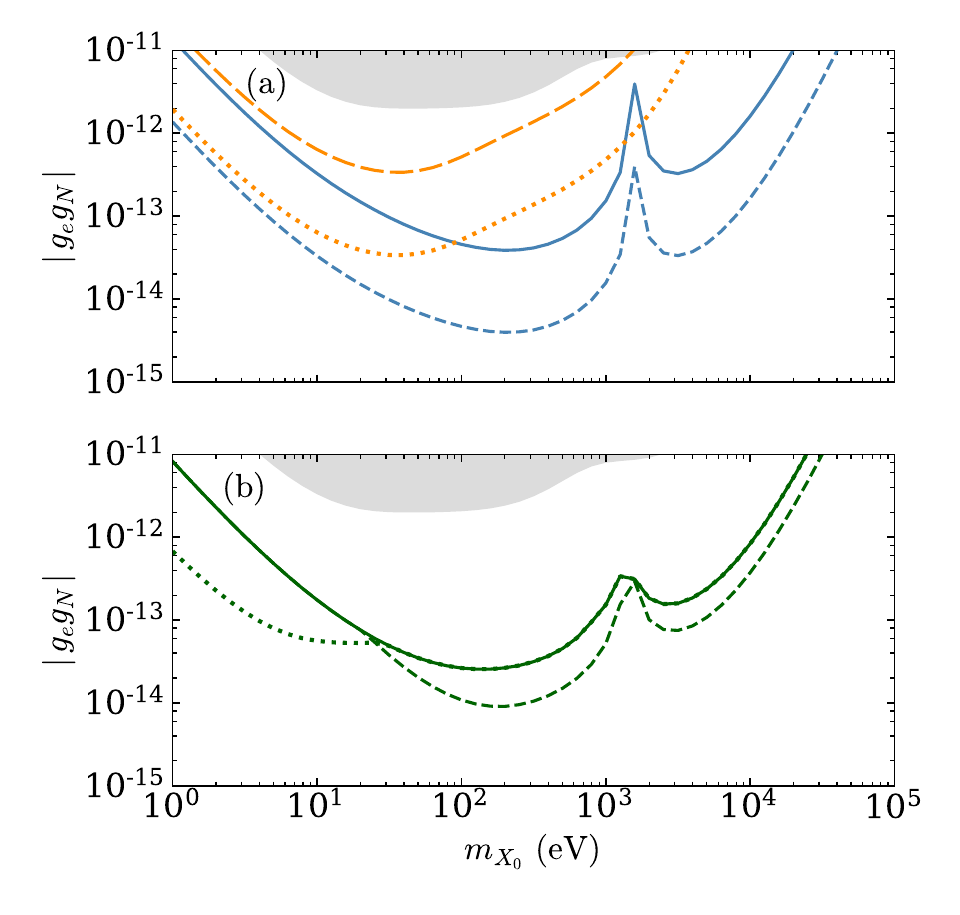}
\caption{Upper bounds on the possible value of $|g_e g_N|$ for an attractive
interaction, as derived
from hypothetical spectroscopic data. For comparison, the region excluded by the analysis of the current data is represented by a shaded area. (a) Solid curve and long-dashed curve: Bounds based on a set of transitions between
s-states (solid curve) and between d-states
(long-dashed curve), assuming a 10~Hz experimental error and a theoretical error scaling as stated in the text. Short-dashed curve and dotted curve: the same as respectively the solid curve and the long-dashed curve but assuming a 1~Hz experimental error.
(b) Solid curve: Bound obtained by comparing the
value of the Rydberg constants derived from the same sets of transitions between s-states and between d-states as in panel (a). Dashed curve and dotted curve: the same as the solid curve but with
a further comparison with values of the
Rydberg constant derived from transitions
between circular states.
}
\label{fig:bounds_paper2}
\end{figure}
Fig.~\ref{fig:bounds_paper2} illustrates the improvement on the NP bounds which could be expected from reducing the experimental error on transition frequencies to the 10~Hz level or to an aspirational 1~Hz level.

Each of the bounds shown in Fig.~\ref{fig:bounds_paper2}(a) was obtained by comparing the predictions of the Standard Model to a set of hypothetical data, the latter having been generated from a model including a NP shift. The details of the calculation are given in
Appendix~\ref{app:fittingproj}.

    The two blue curves plotted in Fig.~\ref{fig:bounds_paper2}(a) represent the bounds derived in this way from an arbitrary and hypothetical set of eight transitions between
    s-states, namely the 1s~--~2s, 2s~--~5s, 2s~--~8s, 2s~--~9s, 2s~--~11s,
    2s~--~15s, 2s~--~21s and 2s~--~30s transitions. As seen from the figure, these results would improve the current spectroscopic bounds by two orders of magnitude over a wide range of values of
    $m_{X_0}$, assuming an experimental error of 10~Hz. Reducing the error to 1~Hz would yield a three orders of magnitude improvement.
    
    Using only transitions between states with $l>0$ would remove the uncertainty on how the theoretical error scales with $n$. In practice, an experimental value for such a transition could be obtained, e.g., by measuring the
    2s to $(n,l)$ and 2s to $(n',l')$ intervals and subtracting one from the other to find the $(n,l)$ to $(n',l')$ interval. The two orange curves plotted in Fig.~\ref{fig:bounds_paper2}(a) represent the bounds derived from a set
    of transitions between d-states only 
    (namely the 8d~--~9d, 8d~--~11d,
    8d~--~15d, 8d~--~21d and 8d~--~30d transitions). While proceeding in this way has the advantage of avoiding the scaling issue, it has the disadvantage of taking into account only states with a relatively small NP shift. Correspondingly, and as is illustrated by the numerical results of Fig.~\ref{fig:bounds_paper2}(a), the bounds derived from such a set of data would be less stringent than those derived from data that include transitions from or between deeply bound states.
    
    As mentioned above, bounds on $g_eg_N$ can also be obtained by comparing the values of Rydberg constants derived from different sets of data. Assuming a 10~Hz experimental error and performing
    this comparison between the same sets of transitions as in Fig.~\ref{fig:bounds_paper2}(a) gives the bound represented by a solid curve in Fig.~\ref{fig:bounds_paper2}(b). This bound is slightly tighter but generally differs little from that obtained directly from the fit of the transitions between s-states. The dashed curve and dotted curve show that this bound could be lowered still further by also comparing these two values of the Rydberg constants with the value derived from transitions between circular states --- i.e., transitions
    of the form $(n,l=n-1) \leftrightarrow (n'=n+1,l'=n'-1)$. We consider two different sets of such transitions in Fig.~\ref{fig:bounds_paper2}(b). We took $n=10$, 15, 20, 25 or 30 and assumed an experimental error of 0.5~kHz on these transitions to
    calculate the bound represented by a dashed curve, whereas for the bound represented by a dotted curve we took $n=40$, 41, 42, 43 or 44 and assumed an experimental
    error of 0.1~Hz. Because the electronic density is concentrated further away from the nucleus when $n > 40$ than when $n \leq 30$, adding the first or the second of these two sets of
    transitions lowers the bound in different ranges of values of $m_{X_0}$.


\section{Summary and Conclusions}
\label{sec:conc}
In summary, we have considered how the entire set of currently available spectroscopic data may be used to set global constraints on NP models that can be parameterized as a Yukawa-type interaction. Such interactions would naturally lead to so-called fifth forces which are a being searched for intensively \cite{Adelberger:2003zx,Berengut:2017zuo}.

Light force mediators have been intensively tested in lab experiments, e.g. through the Casimir effect \cite{Bordag:2001qi}. As such searches rely in general on all atoms in a macroscopic object to contribute coherently and in concert to the resulting force on a test object, they do not probe directly the existence of a force on a microscopic level. This leaves large classes of new physics models untested. For example, forces mediated via kinetic mixing between a photon and a new $Z'$ \cite{Holdom:1985ag,Foot:1991kb} can easily avoid such bounds, as the atom as a whole is not charged under the fifth force. The experiments discussed above, however, would remain sensitive to such an interaction. 

In addition, while other laboratory based experiments lose sensitivity for mediator masses above $100~\mathrm{eV}$, atomic spectroscopy for hydrogen atoms retains a good sensitivity up to masses of $10~\mathrm{keV}$. Thus, to our knowledge, the presented predicted limits provide the strongest constraints in laboratory based experiments obtained so far for that mass range.

The bounds we obtained in this work appear to be weaker than those set by astrophysical bounds. However, astrophysical bounds rely on the thermal production of light force mediators in stars \cite{Grifols:1986fc,Grifols:1988fv}. Particles like chameleons avoid such production and thereby constraints from measurements of the energy transport in stars. Here atomic spectroscopy can help to close gaps in the landscape of Standard Model extensions and provide an independent test of the physics models underlying the assumptions of the models for the evolution of stars.

We further argue that this type of laboratory-based bound is unique, since it is independent of any many-body physics effects, such as astrophysical models or the complex subtleties of isotope shifts in many-electron atoms. Global constraints of this type also reduce the sensitivity to systematic errors in individual measurements, such as those which are currently giving rise to the so-called proton radius puzzle.

We therefore argue that there is a strong case for improved measurements in hydrogen based on extensions of current methods for precision optical frequency measurements in laser-cooled and trapped atoms. An important element would be extending the reach of measurements to higher principal quantum numbers,  which has substantial benefits due to the dependence of the new physics shift on the shape of the wave function discussed in Section~\ref{sec:NPshifts}. The ideal platform would be trapped single atoms or arrays of atoms with a well-controlled spacing, such as an optical tweezer array, opening also the tantalising prospect of an engineered many-body quantum system with a complete SM description.

An extension of this work would consider other simple atoms which have a complete SM description, such as D and He, or even positronium, muonic hydrogen or muonium. More sophisticated statistical analysis methods might enable measurements in all of these systems to be combined into a highly robust extended extended bound, or to create sensitive differential searches. Concrete limits could be obtained for various classes of new physics models, e.g., chameleons or kinetic mixing. 

\acknowledgements
We gratefully acknowledge the contribution of Masters students Andrew Spiers, Natalie McAndrew-D'Souza and Suniyah Minhas to the early stages of this work. We also acknowledge helpful discussions with Martin Bauer, David Carty, Stephen Hogan and Joerg Jaeckel, and thank
Savely Karshenboim and Krzysztof Pachucki for information about QED corrections. This work made use of the Hamilton HPC Service of Durham University. MPAJ acknowledges funding from the EPSRC responsive mode grant EP/R035482/1. The project also received funding from the European Union's Horizon 2020 research and innovation programme under the EMPIR grant agreement 17FUN03-USOQS.

\appendix

\section{\label{app:conversion} Conversion between atomic and natural units}

Eq.~(\ref{eq:yukpot}) being written in natural units,
$m_{X_0}$ is expressed as an energy and $r$ as the inverse of an energy.
In atomic units, we have, instead,
\begin{equation}
V(r') = (-1)^{s+1}\frac{B}{r'}~e^{-Cr'},
\label{eq:Vau}
\end{equation}
where the distance $r'$ is expressed in units of the Bohr radius $a_0$, $C$ in units of $a_0^{-1}$ and 
$B$ in units of $\alpha^2\,m_e\,c^2\,a_0$, with $m_e$ the mass of the electron and $\alpha$ the fine
structure constant.
If $r$ and $r'$ refer to the
same point of space and $r$ is expressed in eV$^{-1}$, then
$r' = r \hbar c$ with the product $\hbar c$ expressed
in units of eV~$a_0$
(the product $\hbar c$ has the physical dimensions
of an energy times a length and has a numerical value of 1 in natural
units).
Moreover, as $m_{X_0}r \equiv Cr'$ if $r$ and $r'$ refer to the
same position, we see that $m_{X_0}$ and $C$ are related by the equation
\begin{equation}
C\,[a_0^{-1}] = {m_{X_0}\,[\mbox{eV}] \over (\hbar c)[\mbox{eV}\,a_0]}.
\end{equation}
That is,
\begin{equation}
C\,[a_0^{-1}] = 2.68172763\times 10^{-4}\,\, m_{X_0}\,[\mbox{eV}].
\end{equation}

To relate the constant $B$ to $g_eg_N/4\pi$, we note that in natural units
$g_eg_N$ is a pure number. However, since $V(r)$ is actually an energy and
$r$ a length, Eq.~(\ref{eq:yukpot}) should really be written as
\begin{equation}
V(r) = (-1)^{s+1} {g_e g_N \over 4 \pi}\, \hbar c \, {\exp(-m_{X_0} r) \over r}.
\end{equation}
Thus $B$, in atomic units, is $(g_e g_N /4\pi) \hbar c$ with 
$\hbar c$ expressed as a multiple of the product $E_{\rm h}\,a_0$. Since $E_{\rm h}\,a_0
=\alpha \hbar c$, $\hbar c = (1/\alpha) E_{\rm h}\,a_0$. This gives, to
10~s.f.,
\begin{align}
B[E_{\rm h}\;a_0] =& (137.0359991/4\pi)\,g_eg_N \nonumber \\
                  =& 10.90497832\,g_eg_N.
\end{align}

\section{The NP shift in closed form}
\label{app:analyticalforms}

Eq.~(\ref{eq:NPshift}) can be integrated analytically for any $n$ and $l$. The result is particularly simple for the states with maximum value of $l$ ($l = n-1$):
in the notation of Eq.~(\ref{eq:Vau}) (we use atomic
units throughout this appendix),
\begin{equation}
\delta E_{nn-1}^{\rm NP} = (-1)^{s+1}\,{B \over n^2}\left({nC \over 2} + 1\right)^{-2n}.
\end{equation}
In particular, the NP shift of the 1s state is obtained as
\begin{equation}
\delta E_{10}^{\rm NP} = (-1)^{s+1}\,{4 B \over (C + 2)^2}.
\end{equation}
Note the fast decrease of $\delta E_{nn-1}^{\rm NP}$ for $n\rightarrow \infty$, which arises
from the suppression of the electronic density at small values of $r$ caused by the increasingly strong
angular momentum potential barrier.

The analytical form of $\delta E_{nl}^{\rm NP}$ becomes rapidly unwieldy when $n-l$ exceeds 2 or 3. We quote
results only for the important cases of the 2s, 3s and 4p states, here:
\begin{align}
\delta_{20}^{\rm NP} &= (-1)^{s+1}\,{B \over 2}\,{C^2 + 1/2 \over (C+1)^4}, \\
\delta_{30}^{\rm NP} &= (-1)^{s+1}\,{4B \over 3}\left[
{1\over (3C+2)^2}
-{8\over (3C+2)^3} \right. \nonumber \\&  \; \; \left.
+{32\over (3C+2)^4}
-{64\over (3C+2)^5}
+{160/3\over (3C+2)^6}\right],\\
\delta_{41}^{\rm NP} &= (-1)^{s+1} {B \over 8}\left[
{5\over (2C+1)^4}
-{20\over (2C+1)^5} \right. \nonumber \\ & \; \; \left.
+{35\over (2C+1)^6}
-{30\over (2C+1)^7}
+{21/2\over (2C+1)^8}\right].
\end{align}

\section{Details of the fitting procedure}
\label{app:fitting}

\subsection{Bounds derived from current data}
\label{app:fittingcurr}
The calculation is outlined in Section \ref{section:current}. The full experimental dataset (set A) consists of the 18 measurements labelled A26.1 to A40.2 in
Ref.~\cite{Codata2017}, supplemented by the results of Refs.~\cite{Beyer2017,Fleurbaey:2018fih} and by a transition frequency for the transition between the $n=27$ and $n=28$ circular states calculated from the value of ${\cal R}$ quoted in Ref.~\cite{DeVries2002}. (This value of ${\cal R}$ was derived from a small set of measurements of the $n=27$ to $n=28$ and $n=29$ to $n=30$ transitions, the former weighting more in the determination of ${\cal R}$ than the latter. No recommended value for either of these two transition frequencies is given in Ref.~\cite{DeVries2002}.) The set of data includes a measurement of the 2s$_{1/2}$~--~2p$_{1/2}$ Lamb shift \cite{Pipkin1981}
recently reanalyzed in \cite{Marsman18}. We use the revised value of this experimental result rather than its original value.

The correlation coefficients between the 18 measurements mentioned in Ref.~\cite{Codata2017} are given in that reference. We take the errors on these measurements to be uncorrelated with the errors on the measurements of Refs.~\cite{Beyer2017,Fleurbaey:2018fih,DeVries2002} and the latter to be uncorrelated with each other.

The calculation of the terms $R_p^2\,\tilde{\Delta}_{b_ia_i}^{\rm ns}$ and $\Delta_{b_ia_i}^{\rm oc}$ follows \cite{Horbatsch2016} in regards to the hyperfine splitting and \cite{Yerokhin2019} in regards
to the Lamb shift. Ref.~\cite{Yerokhin2019} updates and completes the
review of hydrogen theory given in the CODATA
compilations \cite{Codata2010,Codata2014}, in particular by
taking into account a number
of more recent investigations \cite{Yerokhin2015,Yerokhin2016,Czarnecki2016,Yerokhin2018,Pachucki2018,Karshenboim2018,Tomalak2019}.
We include all the corrections
listed in \cite{Yerokhin2019} (the details of the original
publications can be found in this reference). However, 
we use the result of \cite{Karshenboim2019a,Karshenboim2019b} for the light-by-light contribution, set the $C_{50}$ coefficient to
the value found in \cite{Karshenboim2019a,Karshenboim2019c}, and use the value
recommended by \cite{Karshenboim2019d} for the pure self-energy
two-loop remainder term. Doing so reduces
the theoretical uncertainties to 1.2~kHz for the ground state energy and to (1.8~kHz)$/n^3$ for the excited states energies (excluding the contribution from
the uncertainty on $R_p$). We treat the resulting errors on the terms
${\Delta}_{b_ia_i}^{\rm oc}$ as completely correlated.
We neglect the theoretical uncertainty on the energies of the states with $l > 0$, which is considerably smaller. The theoretical error on the factors 
$\tilde{\Delta}_{b_ia_i}^{\rm g}$ and $\tilde{\Delta}_{b_ia_i}^{\rm ns}$ is also too small to be relevant in the present context.

\subsection{Projected bounds}
\label{app:fittingproj}
We assume that
each of the measured transition frequencies included in the set can be written
as a sum of the form
$${\cal R}_{0{\rm H}}\left(\frac{1}{n_a^2}-\frac{1}{n_b^2}\right) + \Delta_{ba}^{\rm corr} +
    \Delta_{ba}^{\rm NP},$$
    within experimental error,
    where ${\cal R}_{0{\rm H}} = {\cal R}_0(m_{\rm r}/m_e)$ with
    ${\cal R}_0$ the true value of the Rydberg frequency and
    $\Delta_{ba}^{\rm corr}$ is the sum of all the QED, hyperfine and other corrections predicted by the Standard Model. We take ${\cal R}_{0{\rm H}}$ and
    $\Delta_{ba}^{\rm corr}$ to be the exact values of these quantities. We equate each of the experimental intervals to its Standard Model prediction, ${\cal R}_{{\rm H}}({1}/{n_a^2}-{1}/{n_b^2}) + \Delta_{ba}^{\rm corr} + \alpha_{ba}^{\rm corr}$, where ${\cal R}_{\rm H}$ is an effective Rydberg frequency obtained by fitting theory to experiment and $\alpha_{ba}^{\rm corr}$ represents the theoretical error on $\Delta_{ba}^{\rm corr}$. (This last term thus accounts for the error introduced by the uncertainty on the
    value of $R_p$ as well as the errors on the values of the QED and other corrections not calculated to a sufficient precision. We do not need to consider the uncertainty on the mass ratio $(m_{\rm r}/m_e)$ separately from the uncertainty on ${\cal R}$ since this ratio is subsumed into the fitting parameter ${\cal R}_{\rm H}$.)
    We assume that the $n^{-3}$ scaling mentioned in Section \ref{section:improvedtheory} holds for the s-states of interest, and that the theoretical error on the states with $l>0$ is negligible Doing so for each of the $N$ transitions of a same set yields the following overdetermined system:
    \begin{align}
        &{\cal R}_{{\rm H}}\left(\frac{1}{n_{a_i}^2}-\frac{1}{n_{b_i}^2}\right) + \Delta_{{b_i}{a_i}}^{\rm corr} + A\left(\frac{\delta_{l_{b_i}0}}{n_{b_i}^3}-\frac{\delta_{l_{a_i}0}}{n_{a_i}^3}\right) = \qquad \nonumber \\
        & \qquad \quad  {\cal R}_{0{\rm H}}\left(\frac{1}{n_{a_i}^2}-\frac{1}{n_{b_i}^2}\right) + \Delta_{{b_i}{a_i}}^{\rm corr} +
    \Delta_{{b_i}{a_i}}^{\rm NP} \pm \alpha_{{b_i}{a_i}}^{\rm exp}, \nonumber\\
    &\qquad\qquad \qquad \qquad \qquad \qquad \qquad \quad
    n = 1,2,\ldots,N,
    \end{align}
    where $A$ is a constant,
    and $\alpha_{{b_i}{a_i}}^{\rm exp}$ represents the experimental error
    on the corresponding transition frequency. Simplifying these equations gives
    \begin{align}
        &\delta{\cal R}_{{\rm H}}\left(\frac{1}{n_{a_i}^2}-\frac{1}{n_{b_i}^2}\right) + A\left(\frac{\delta_{l_{b_i}0}}{n_{b_i}^3}-\frac{\delta_{l_{a_i}0}}{n_{a_i}^3}\right) = \qquad \qquad \nonumber \\
        & \qquad \qquad \qquad  
    \Delta_{{b_i}{a_i}}^{\rm NP} \pm \alpha_{{b_i}{a_i}}^{\rm exp}, 
    \qquad
    n = 1,2,\ldots,N,
    \end{align}
    with $\delta{\cal R}_{\rm H} ={\cal R}_{\rm H}-{\cal R}_{0{\rm H}}$. We determine the NP bounds by finding the range of values of $g_eg_N$ within which the left-hand sides of these equations fit the right-hand sides at the 
    5\% confidence level, treating $\delta{\cal R}_{\rm H}$ and $A$ as fitting parameters.  As the experimental errors on different transitions measured using a same methodology
    could be expected to be mildly correlated, we assume a correlation coefficient of 0.1 between the experimental errors on the measured transition frequencies belonging to a same set of data.
    
    The bounds shown in Fig,~\ref{fig:bounds_paper2}(b) follow from Eq.~(\ref{eq:bounds2}), with the Rydberg frequencies ${\cal R}_C$ and ${\cal R}_D$ replaced by the corresponding values of $\delta {\cal R}_{\rm H}$. For simplicity, we assume no correlation between the errors on these quantities.

\section{Black body radiation}
\label{BBR}
We have calculated
the BBR shift of the Rydberg states of interest
in order to ascertain
the precision on the thermometric measurements required in our approach.
Our results complete those of Refs.~\cite{FarnleyWing} and
\cite{Solovyev15}, which do not extend high enough in principal quantum
numbers. Except where specified otherwise, we use atomic units throughout this appendix. 

To second order in the electric field component of the BBR field, the 
thermal
shift of a state $a$ at a temperature $T$ can be written as \cite{FarnleyWing,Yorke83,Solovyev15}
\begin{equation}
\delta_a^{\rm BB} = {2(k_{}T)^3 \over 3 \pi c^3} \sum_{i,b}
|\left< b | r_i | a \right>|^2 F_{}\left({\omega_{ab}\over k_{}T}\right),
\label{eq:shiftBBR}
\end{equation}
where the $r_i$'s are three orthogonal components of the electron's position
operator, $k$ is Boltzmann constant, the summation over
$b$ runs over all the atomic states dipole-coupled to the state $a$, $\omega_{ab} = E_a-E_b$ where
$E_a$ and $E_b$ are the energies of the respective states, and
\begin{equation}
F_{}(y) = {\rm P.V.}\,\int_0^\infty{2y \over y^2-x^2}\, {x^3 \over e^x -1}\,{d}x.
\end{equation}
The BBR field also depopulates state $a$
by inducing transitions to
other states at a rate approximately equal to $\Gamma_a^{\rm BB}$, where
\begin{equation}
\Gamma_a^{\rm BB} = {4(k_{}T)^3 \over 3 c^3} \sum_{i,b}
|\left< b | r_i | a \right>|^2\, U_{}\left({\omega_{ab}\over k_{}T}\right),
\label{eq:GammaBBR}
\end{equation}
with $U(y) = |y|^3/(\exp|y|-1)$ \cite{FarnleyWing,Yorke83,Solovyev15}.
$\Gamma_a^{\rm BB}$ does not include losses due to the
BBR-induced Stark mixing of degenerate states of opposite parity, which
is significant in hydrogen \cite{Solovyev15}. Eqs.~(\ref{eq:shiftBBR})
and (\ref{eq:GammaBBR}) also neglect
non-dipolar transitions and corrections
of fourth order in the BBR electric field
\cite{Palchikov03,Porsev06,Safronova13}; their contributions are small and can be neglected for our purposes. 
Local anisotropies of the BBR field may also need to be factored in
when comparing to experiment \cite{Flambaum16}.

We evaluate $F(y)$ by contour integration in the complex $x$-plane.
This method bypasses the need of a
careful treatment of the singularity at $x=\pm y$
inherent in the direct calculation of a Cauchy principal value
and only involves straightforward numerical quadratures.
Namely, we introduce the complex function
\begin{equation}
F_C(y) = \int_C {2y \over y^2-z^2}\, {z^3 \over e^z -1}\,{d}z,
\end{equation}
where the integration contour starts at $z=0$ and goes to $\mbox{Re}\,z \rightarrow \infty$ in 
the lower half plane, avoiding the zeroes of
$\exp(z)-1$. In practice, we use a rectangular contour running from $z=0$ to $z=-i/2$ and from $z=-i/2$ to $z=50-i/2$, which is well adapted to the range of values of
$y$ involved in this work. We have $F_{}(y) \equiv \mbox{Re}\,F_C(y)$ owing to
the relation
\begin{equation}
\lim_{\epsilon \rightarrow 0^+}\,{ 1 \over y - x +i\epsilon} =
{\rm P.V.}\,{1 \over y - x}
-i\pi\delta(y-x), 
\end{equation}
and moreover
\begin{equation}
\delta_a^{\rm BB} - {i \over 2} \Gamma_a^{\rm BB} =
{2 (k_{}T)^3 \over 3 \pi c^3} \sum_{i,b}
|\left< b | r_i | a \right>|^2 F_{C}\left({\omega_{ab}\over k_{}T}\right).
\end{equation}
We calculate the dipole matrix elements $\left< b | r_i | a \right>$ by solving
Eq.~(\ref{eq:eigenvalueprob}) for each of the states
$a$ and $b$. Having the corresponding generalized eigenvectors,
${\sf c}_a$ and ${\sf c}_b$, we obtain $\left< b | r_i | a \right>$
as ${\sf c}_b^\dagger {\sf R}_i {\sf c}_a$, where ${\sf R}_i$ is the matrix
of elements
\begin{equation*}
\int S_{n'l_b}^*(r) Y_{l_b m_b}^*(\theta,\phi) \, r_i \,
S_{n'l_a}(r) Y_{l_a m_a}(\theta,\phi) \, {d}^3r.
\end{equation*}
Substituting these results into Eq.~(\ref{eq:shiftBBR}) gives the shift in the
non-relativistic approximation. We correct this for spin-orbit coupling by replacing the non-relativistic angular factors by the appropriate expressions
\cite{FarnleyWing} and evaluating the Bohr transition frequencies $\omega_{ba}$ using the relativistic energies. 
The summation over the intermediate states $b$ runs over all the
generalized eigenvectors of the matrix $\sf H_0$ of the relevant symmetry,
including those corresponding to positive eigenenergies. Doing so 
ensures (assuming that the basis is large enough) that the shifts and widths properly include the
contribution of the continuum, which can be significant
\cite{Glukhov10,Ovsiannikov12}. We use 300 Sturmian functions for
each symmetry. While the BBR shift depends
to some extent on the
hyperfine structure of the levels \cite{Itano82},
taking it into account 
would not affect
the results at the level of precision required by the present
investigation. The calculations of radiative widths mentioned in the text use exactly the same numerical method in regards
to the computation of the required dipole matrix elements.

\begin{figure}[t]
\centering
\includegraphics[width=8.5cm]{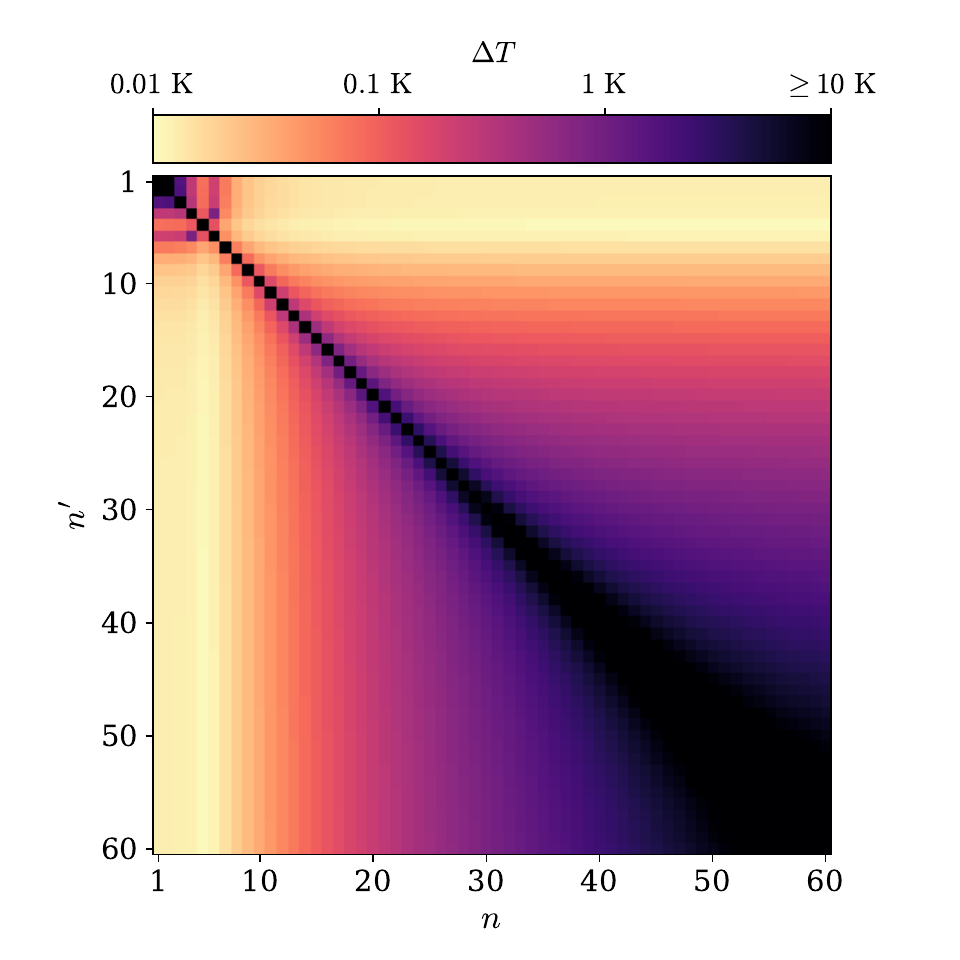}
\caption{
Difference between 300~K and the temperature at which
the frequency of the transition from
the $(n,l=0)$ state
to the $(n',l=0)$ state differs by 0.2~Hz from its
value at 300~K.
}
\label{fig:maxtempdiff}
\end{figure}
The BBR shift of hydrogen state may be significant compared to the
NP shift at the relevant values of $g_eg_N$, and may even be considerably larger. For example,
for $n=10$ and $l=0$, $|\delta_{nl}^{\rm NP}|$ is at most 0.7~kHz when $g_eg_N = 1\times 10^{-12}$ whereas
$\delta_{nl}^{\rm BB}$ is approximately 1.1~kHz at 300~K \cite{FarnleyWing}.
At least in principle, this shift can be removed from
spectroscopic data for hydrogen since it can be accurately calculated
for this atom. In practice, however, taking it correctly into account 
requires a sufficiently precise determination of
the temperature of the BBR field at the location of the atoms, and perhaps also
of its inhomogeneity and its deviation from of an ideal
Planck distribution. {\it In situ} temperature measurements 
with an uncertainty of the order of 0.01~K have been achieved
using platinum resistance thermometers \cite{Nicholson15}.
Spectroscopic measurements of Rydberg states have also been proposed to
determine the temperature of the BBR background with a similar uncertainty
\cite{Ovsiannikov11}.

The BBR energy shift of the high Rydberg states
is approximately $\pi(kT)^2/3c^3$  \cite{GallagherCooke,FarnleyWing}.
An error of 0.01~K on $T$ at 300~K translates into an error of
0.2~Hz (or lower) on the BBR shift of these states. However,
this error on the temperature would have a smaller impact
on measurements of the energy difference between Rydberg states
made in a same apparatus because they all shift by roughly the same
amount.
This point is illustrated by Fig.~\ref{fig:maxtempdiff}, which shows the accuracy to which $T$ must be known to reduce the error on the BBR shift to less than 0.2~kHz in measurements of transitions between s-states made at room temperature. An easily achievable accuracy of 0.5~K is sufficient for transitions between the lowest states or between high Rydberg states (the former because they shift little, the latter because they shift similarly). The requirements are more stringent for transitions between relatively low lying states and high Rydberg states, particularly for low lying states with $n \approx 5$ (whose shift is larger and of opposite sign to that of states with $n \gg 5$ \cite{FarnleyWing}). 

It should be noted that the error that can be tolerated on $T$ roughly scales with the maximum error on the transition frequencies. If an accuracy of 10~Hz is sought, rather than 0.2~Hz, knowing the temperature would not need to be known to better than 0.5~K for any frequency interval.


\bibliography{allreferences_resubmitted}
\end{document}